\documentstyle[12pt]{article}

\def\chain#1#2{\mathrel{\mathop{\null\longrightarrow}\limits^{#1}_{#2}}}

\def\title#1#2#3#4#5{\thispagestyle{empty}
        \begin{center} \begin{tabular}[t]{l} #1 \end{tabular}
        \hfill	\begin{tabular}[t]{r} #2 \end{tabular}
        \\[1cm] {\LARGE\bf #3} \\[.5in] {#4{}}
        \end{center} \vfill \centerline{{\large ABSTRACT}}
	   {\nopagebreak \noindent\begin{quotation}\noindent {\small #5}
	\end{quotation}} \vfill \newpage
        \def\thefootnote{\sharp\arabic{footnote}}}

\begin {document}
\title {July 1992} {OITS-488}
{Implications of LEP results for SO(10)
grandunification  with two intermediate stages}  {
{\bf N.G. Deshpande, E. Keith, Palash B. Pal}\\  {\em Institute of
Theoretical Science\\ University of Oregon, Eugene, OR
97403-5203}}   {We consider the breaking of the grand unification
group $SO(10)$ to the standard model gauge group through several
chains containing two intermediate stages.  Using the values of the
gauge coupling constants at scale $M_Z$ derived from recent LEP
data, we determine the range of their intermediate and unification
scales.  In particular, we identify those chains that permit new
gauge structure at relatively low energy $(\sim 1\, {\rm TeV})$.}

Recently, $SO(10)$ \cite{so10} breaking chains  with
one-intermediate stage \cite{1stage} have been examined in light of
the latest LEP data \cite{lep}.  This data gives
   \begin{eqnarray}
\alpha _1 (M_Z) &=& 0.016887 \pm 0.000040\nonumber\\
\alpha _2 (M_Z) &=& 0.03322 \pm 0.00025\nonumber\\
\alpha _3 (M_Z) &=& 0.120 \pm 0.007
   \end{eqnarray}
where the $\alpha_i$s are normalized such that they would be
equal when $SO(10)$ is a good symmetry and refer to $U(1)_Y$,
$SU(2)_L$ and $SU(3)_c$ respectively.  Our conclusion was that if
$SO(10)$ breaks through a single intermediate scale to the standard
model, then this scale is in the range of $10^9$ to  $10^{11}$ GeV.
In this report, we extend our analysis to two intermediate stage
breaking schemes. Such analysis has been done previously
\cite{previous}. Our analysis differs from these in the use of the
most recent data given above. We are primarily interested in
identifying those chains that permit low energy gauge groups
containing the standard model as a subgroup.  We find that it is
possible to have extra neutral gauge bosons in the low energy
regime, but definitely no extra charged ones below about $10^7$ GeV.

We start by noting that all  possibilities with grand unified
$SU(5)$ in the intermediate stage are already ruled out by the
data. So we  look at symmetry breaking chains where the
intermediate level gauge groups  are either $\{2_L2_R4_CP\}$ or any
of its subgroups, where $2_L$, for example, stands for the group
$SU(2)_L$  and $P$ denotes an unbroken $L\leftrightarrow R$ parity
symmetry. All such chains are listed below, where we have also
indicated the representation of Higgs multiplet responsible for
breaking at each stage.
  \begin{eqnarray} {\rm I:}  &&
SO(10) \chain{}{210}  \{2_L 2_R 4_C\}
\chain{}{45}  \{2_L 2_R 1_X 3_c\}
\chain{}{h}   \{2_L 1_Y 3_c\}  \nonumber
\\
 {\rm II:} &&
SO(10) \chain{}{54}  \{2_L 2_R 4_C P\}
\chain{}{ 210 }   \{2_L 2_R 1_X 3_c P\}
\chain{}{h}   \{2_L 1_Y 3_c\} \nonumber
\\ {\rm III:}  &&
SO(10) \chain{}{ 54 }  \{2_L 2_R 4_C P\}
\chain{}{ 45 }  \{2_L 2_R 1_X 3_c\}
\chain{}{h}   \{2_L 1_Y 3_c\} \nonumber
\\ {\rm IV:}  &&
SO(10) \chain{}{ 54 }  \{2_L 2_R 1_X 3_c P\}
\chain{}{ 210 }  \{2_L 2_R 1_X 3_c\}
\chain{}{h}   \{2_L 1_Y 3_c\} \nonumber
\\  {\rm V:}  &&
SO(10) \chain{}{ 210 }  \{2_L 2_R 4_C\}
\chain{}{45}  \{2_L 1_R 4_C\}
\chain{}{h}   \{2_L 1_Y 3_c\} \nonumber
\\ {\rm VI:}  &&
SO(10) \chain{}{ 54 }  \{2_L 2_R 4_C P\}
\chain{}{ 45 }  \{2_L 1_R 4_C\}
\chain{}{h}   \{2_L 1_Y 3_c\} \nonumber
\\ {\rm VII:}  &&
SO(10) \chain{}{ 54 }  \{2_L 2_R  4_C P\}
\chain{}{ 210 }  \{2_L 2_R 4_C\}
\chain{}{h}   \{2_L 1_Y 3_c\} \nonumber
\\ {\rm VIII:}  &&
SO(10) \chain{}{ 45 }  \{2_L 2_R 1_X 3_c\}
\chain{}{ 45 }   \{2_L 1_R 1_X 3_c\}
\chain{}{h}   \{2_L 1_Y 3_c\} \nonumber
\\ {\rm IX:}  &&
SO(10) \chain{}{ 54 }  \{2_L 2_R 1_X 3_c P\}
\chain{}{ 45 }  \{2_L 1_R 1_X 3_c\}
\chain{}{h}   \{2_L 1_Y 3_c\} \nonumber
\\ {\rm X:}  &&
SO(10) \chain{}{ 210 }  \{2_L 2_R 4_C\}
\chain{}{  210 } \{2_L 1_R 1_X  3_c\}
\chain{}{h}   \{2_L 1_Y 3_c\} \nonumber
\\  {\rm XI:}  &&
SO(10) \chain{}{ 54 }  \{2_L 2_R 4_C P\}
\chain{}{ 210 }  \{2_L 1_R 1_X 3_c\}
\chain{}{h}   \{2_L 1_Y 3_c\} \nonumber
\\ {\rm XII:}  &&
SO(10) \chain{}{ 45 }  \{2_L 1_R 4_C\}
\chain{}{ 45 }  \{2_L 1_R 1_X 3_c\}
\chain{}{h}   \{2_L 1_Y 3_c\} \label{chains}
  \end{eqnarray}
In the above equations, $X = {B-L \over 2}$. In all cases, $\{2_L
1_Y 3_c\}$ breaks to $\{3_c 1_Q\}$ with a complex $10$. The
breaking to the standard model is done with $h$, where $h$ can be a
$16$ or $126$ dimensional representation of $SO(10)$. In either
case,  we can achieve see-saw mechanism to generate small neutrino
masses,  at the tree level \cite{ygrs} with 126, or  through loops
\cite{witten} using a 16.

\begin{table}
\begin{tabular}{|c|l|} \hline\hline
Intermediate gauge group & Higgs contribution $T(S_i)$ \\
\hline  $\{2_L1_R 4_C\}$ & $T_{2L}=1\phi^{10}$\\
             & $T_{1R} = 1\phi^{10} + 20 \Delta^{126}_R + 2\delta^{16}_R$ \\
													& $T_{4C} = 6 \Delta^{126}_R + 1 \delta^{16}_R +
																4\Lambda^{45} +4\Lambda^{210}$\\ \hline
$\{2_L 2_R 4_C\}$ & $T_{2L} = 2 \phi^{10} + 40 \Delta^{126}_L
+ 4\delta^{16}_L + 2\Sigma^{45}_L + 30\sigma^{210}_L$\\
              & $T_{2R} = 2\phi^{16} + 40
\Delta^{126}_R + 4 \delta^{16}_R + 2 \Sigma^{45}_R +30\sigma^{210}_R$\\
              & $T_{4C} = 18 \Delta^{126}_R + 18
\Delta^{126}_L + 2 \delta^{16}_R + 2 \delta^{16}_L  + 12
\sigma^{210}_R + 12 \sigma^{210}_L$\\
           & $\quad + 4 \Lambda^{45} + 4\Lambda^{210}$\\ \hline
$\{2_L 2_R 1_X 3_c\}$ & $T_{2L} = 2\phi^{10} + 4 \Delta^{126}_L
+ 1\delta^{16}_L + 2\Sigma^{45}_L$\\
                  & $T_{2R} = 2\phi^{10}
+ 4 \Delta^{126}_R + 1\delta^{16}_R + 2\Sigma^{45}_R $\\
                  & $T_{1X} = 9 \Delta^{126}_L + 9 \Delta^{126}_R +
{3\over2}\delta^{16}_L + {3\over2}\delta^{16}_R$\\
                  & $T_{3c} = 0$\\ \hline
$\{2_L 1_R 1_X 3_c\}$ & $T_{2L} = 1\phi^{10}$\\
                  & $T_{1R} = 1\phi^{10} + 2 \Delta^{126}_R + {1 \over 2}
\delta^{16}_R$  \\
                  & $T_{1X} = 3 \Delta^{126}_R + {3 \over 4} \delta^{16}_R$\\
                  & $T_{3c} = 0$\\ \hline\hline
\end{tabular}
\caption{Expressions for $T(S_i)$ for different intermediate gauge
groups.} \end{table}
                To examine these cases, we use the one-loop
renormalization group equations (RGE's)
  \begin{eqnarray}
\mu {\partial \alpha _i \over \partial \mu} = {b_i \over 2 \pi}
\alpha^2_i \, ,
  \end{eqnarray}
which gives
\begin{eqnarray}
	\alpha_i^{-1}(M_2)=\alpha_i^{-1}(M_1)-{b_i\over2\pi} \ln {M_2 \over M_1}\,   .
 \end{eqnarray}
Here,
   \begin{eqnarray}
b_i = {4 \over 3} n_g - {11 \over 3} N + {T(S_i) \over 6}
   \end{eqnarray}
where $n_g$ is the number of fermion generations which we take as
three and $N$ is the value of $N$ in $SU(N)$  with $N = 0$ for
$U(1)$.  The necessary expressions for $T(S_i)$ are given in Table
1.  If the Higgs fields are complex then the value of $T(S_i)$ has
been multiplied by a factor of two. We use the hypothesis  of
minimal fine-tuning \cite{finetune}, which fixes the masses of Higgs
bosons according to their transformation under the unbroken
subgroups at any scale.  The Higgs fields that enter Table 1 are
those submultiples that have masses below the energy level of
interest and contribute to evolution of the couplings. (The Greek
symbols take on the values of the numbers of corresponding
submultiplets with masses less than the scale of interest.)  These
submultiplets are defined in Table 2.

\begin{table}
\begin{tabular}{|c|c|c|c|c|c|} \hline
 &\multicolumn{4}{|c|}{Relevant submultiplet in
$SO(10)$ subgroup}& \\ \cline{2-5}
\multicolumn{1}{|c|}{SO(10) multiplet}& $\{2_L
1_R4_C\}$ &$ \{2_L
2_R 4_C\}$ & $\{2_L 2_R
1_X 3_c\}$ & $\{2_L 1_R
1_X 3_c\}$ & Notation\\
\hline
10 & $(2, {1 \over 2}, 1)$ & $(2,2,1)$ & $(2,2,0,1)$ &
$(2,{1 \over 2},0,1)$ & $\phi^{16}$\\
   16 & $(1, -{1 \over 2}, 4)$ & $(1,2,4)$ & $(1,2,{1 \over
2},1)$ & $(1,-{1 \over 2},{1 \over 2},1)$ & $\delta^{16}_R$\\
   16 &  & $(2,1,4)$ & $(2,1,-{1 \over 2},1)$ &  &
$\delta^{16}_L$\\
126 &$(1,1,10)$  & $(1,3,10)$ & $(1,3,-1,1)$ & $(1,1,-1,1)$ &
$\Delta^{126}_R$\\
126 &  & $(3,1,10)$ & $(3,1,1,1)$ &&
$\Delta^{126}_L$\\
45 & $(1,0,15)$ & $(1,1,15)$ &  & & $\Lambda^{45}$\\
210 &  & $(1,1,15)$ &  & & $\Lambda^{210}$\\
45 &  & $(1,3,1)$ &$(1,3,0,1)$  & & $\Sigma^{45}_R$\\
45 &  & $(3,1,1)$ &$(3,1,0,1)$  & & $\Sigma^{45}_L$\\
210 &  & $(1,3,15)$ &  & & $\sigma^{210}_R$\\
210 &  & $(3,1,15)$ &  & & $\sigma^{210}_L$\\ \hline
\end{tabular}
\caption{The submultiplet, whose representation
under the subgroup is shown, contributes to the evolution of the
gauge coupling of that subgroup.} \end{table}
                  In our analysis, we match couplings at each stage
of symmetry breaking.  We assume that all fermions have masses less
than $M_Z$.  Although the mass of the $t$-quark is expected to be
slightly larger than $M_Z$, and the mass of $\nu _R$ could be much
larger than $M_Z$, the corrections due to these are negligible
for the purposes of our calculations.

\begin{table}
\begin{center}
\begin{tabular}{|l|rl|r|} \hline\hline
Chain & \multicolumn{3}{c|}{Allowed values of $n_1$}  \\
\cline{2-4}   & \multicolumn{2}{c|}{Lowest} & Highest \\ \hline
Ia & 8.2 & $\pm 0.2$ & $10.6\pm 0.2$ \\
Ib & 10.0 &$\pm 0.2$ & $13.5\pm 0.2$ \\
IIa & 8.6& $\pm  0.2$ & $13.6\pm 0.2$ \\
IIb & 10.0& $\pm  0.2$ & $13.6\pm 0.2$ \\
IIIa & 8.0& $\pm  0.4$ & $13.6\pm 0.2$ \\
IIIb & 9.8& $\pm  0.2$ & $13.6\pm 0.2$ \\
IVa & 8.2& $\pm  0.2$ & $10.8\pm 0.3$ \\
IVb & 9.8& $\pm  0.2$ & $12.3\pm 0.3$ \\
Va & 11.0& $\pm  0.2$ & $11.2\pm 0.2$ \\
Vb & 12.2& $\pm  0.2$ & $13.6\pm 0.2$ \\
VIa & 11.2& $\pm  0.1$ & $13.8\pm 0.2$ \\
VIb & 12.3& $\pm  0.1$ & $13.6\pm 0.2$ \\
VIIa & 11.3& $\pm  0.2$ & $13.6\pm 0.2$ \\
VIIb & 13.6& $\pm  0.2$ & $13.8\pm 0.2$ \\
VIIIa & $2.0$ && $7.7\pm 0.1$ \\
VIIIb & $2.0$ && $7.5\pm 0.1$ \\
IXa & $2.0$ && $10.0\pm 0.2$ \\
IXb & $2.0$ && $10.6\pm 0.2$ \\
Xb & $2.0$ && $12.2\pm 0.2$ \\
XIa & $2.0$ && $13.5\pm 0.2$ \\
XIb & $2.0$ && $13.6\pm 0.2$ \\
XIIa & $2.0$ && $5.3\pm 0.1$ \\
XIIb & $2.0$ && $12.1\pm 0.2$ \\
\hline\hline \end{tabular}
\end{center}
\caption[]{Acceptable domains of $n_1$ for all chains. The
chains are defined in Eq. (\ref{chains}),  and $a$, $b$ refer to the
choice of $h$  being 126 or 16 respectively.} \end{table}
                   In the study of each chain, we solve analytically
in one loop order for allowed scales, $n_G = \log_{10} \left({M_G
\over GeV }\right)$ at grand unification, $n_2 = \log_{10}
\left({M_2 \over GeV}\right)$, the higher of the intermediate
scales, and $n_1 = \log_{10} \left({M_1 \over GeV }\right)$, the
lower of the intermediate scales.  The graphs for each case
considered are drawn with $1 \sigma$ errors from LEP data.  Only
those portions of the graphs are meaningful  where $n_G \geq n_2
\geq n_1$.  Further $n_G$ has to be sufficiently  high to escape the
constraint from non-observation of proton decay.  We  take this
constraint to be   approximately \cite{1stage}
                 \begin{eqnarray}
\left({\alpha ^{-1} _G \left( M_G \right) \over 40} \right)
\left({ M_G  \over 10^{15} GeV} \right)^2 > 2.5
\label{pdecay}
\end{eqnarray}

Therefore, we consider the portion of any chain where $ M_G <
10^{15} GeV$ to be unacceptable.   We show our findings
in the graphs of Fig. 1, and  in Table 3 we present the acceptable
regions of the lower intermediate scale $n_1$ for these graphs. For
each chain we refer to case (a) where $h=126$ and case (b) where
$h=16$.  Chain Xa is not featured because it has no meaningful
solution.

We find no chain where extra light charged bosons can occur. In all
chains with $SU(2)_R$ or $SU(4)_C$ in the lower intermediate scale,
the allowed regions of $n_1$ tend to be small with both intermediate
scales at very high values. We find that  additional gauge bosons in
the range of TeV's are permitted only in the chains VIII through XII
except for chain Xa which has no meaningful solution. All  of these
chains have $\{2_L 1_R 1_X 3_c\}$ as the lower intermediate gauge
group. This means that there is only one extra gauge boson in the
TeV range, and this one is neutral. Even in this set, the chains XIa
and XIIb are of very marginal acceptability due to the constraint of
Eq. (\ref{pdecay}) set by experiments on proton decay.

In see-saw models of neutrino mass, ordinary neutrinos are light
due to a large Majorana mass of the right-handed neutrinos
\cite{ygrs}. This large Majorana mass cannot be generated unless
the $\{1_R\}$ symmetry is broken. Thus the magnitude of this
Majorana mass is expected to   be similar to the scale of the
$\{1_R\}$ breaking. Our conclusions stated above thus show that it
is possible to obtain this Majorana mass in the TeV range. However,
the full parity symmetry is not restored until at much higher
energy since $\{2_R\}$-breaking always occurs at high scale.

This work has been supported by the Department of Energy grant
DE-FG06-85ER-40224.

\newpage

\centerline{{\large\bf Figure  captions}}

\begin{itemize}

\item[Fig. 1~:]
{{$n_G = \log_{10} \left({M_G \over GeV }\right)$ and $n_2 =
\log_{10} \left({M_2 \over GeV}\right)$ plotted vs. $n_1 = \log_{10}
\left({M_1 \over GeV }\right)$ for chains I through XII. For each
chain we refer to case (a) where $h=126$ and case (b) where $h=16$.
The constraint $n_G,n_2\geq n_1$ is violated by $n_G$ or $n_2$ being
in the shaded portion of the graphs. The acceptable domain for $n_1$
in each case is given in Table 3.} \label{fig1}}

\end{itemize}

\begin{thebibliography}{[001]}

\bibitem{so10} H Georgi: in {\sl Particles and Fields}, ed. by C E
Carlson (American Institute of Physics, 1975); H Fritzsch, P
Minkowski: Ann. Phys. 93 (1975) 193.

\bibitem{1stage} N. G. Deshpande, E. Keith, P. B. Pal: Phys. Rev. D
(to be published).

\bibitem{lep} The values of $\alpha_1$ and $\alpha_2$ have been
taken from  U Amaldi, W de Boer, H F\"{u}rstenau: Phys. Lett. B260
 (1991) 447, and that of $\alpha_3$ from T Hebbeker: plenary talk
at the Lepton-Photon symposium, Geneva, July 1991.

\bibitem{previous}
D Chang, R N Mohapatra,  M K Parida: Phys.
Rev.  D30 (1984) 1052; J M Gipson, R E Marshak: Phys. Rev. D31
(1985) 1705;
D Chang, R N Mohapatra, J M Gipson, R E Marshak, M K Parida: Phys.
Rev.  D31 (1985) 1718.

\bibitem{ygrs} M Gell-Mann, P Ramond, R Slansky: in {\sl
Supergravity}, ed. P van Niewenhuizen and D Z Freedman (North
Holland 1979); T Yanagida: in {\sl Proceedings of Workshop on
Unified Theory and Baryon Number in the Universe}, ed. O sawada
and A Sugamoto (KEK 1979).

\bibitem{witten} E Witten: Phys. Lett. 91B (1980) 81.

\bibitem{finetune} F del Aguila, L Iba\~{n}ez: Nucl. Phys. B177
(1981)  60; R N Mohapatra, G Senjanovi\'{c}: Phys. Rev. D27 (1983)
1601.

\end{thebibliography}
\end{document}